\documentclass[preprints,article,submit,moreauthors,pdftex]{Definitions/mdpi} 
\usepackage{luatex85}

\firstpage{1} 
\pubvolume{1}
\issuenum{1}
\articlenumber{0}
\pubyear{2022}
\copyrightyear{2022}
\datereceived{} 
\dateaccepted{} 
\datepublished{} 
\hreflink{https://doi.org/} 
\Title{The Formulation of Scaling Expansion in an Euler-Poisson Dark-fluid Model}

\TitleCitation{The Formulation of Scaling Expansion in an Euler-Poisson Dark-fluid Model}

\Author{Balázs Endre Szigeti $^{1,2}$\orcidA{}, Imre Ferenc Barna $^{1,\dagger}$\orcidB{} and Gergely Gábor Barnaföldi $^{1,\ddagger}$\orcidC{} }

\AuthorNames{Balázs E. Szigeti, Imre F. Barna and Gergely G. Barnaföldi}

\AuthorCitation{Szigeti, B.; Barna, I.; Barnaföldi, G.}

\address{%
$^{1}$ \quad Wigner Research Centre for Physics, Institute for Particle and Nuclear Physics, Budapest HU-1121\\
$^{2}$ \quad Eötvös Loránd University, Faculty of Science, Department of Atomic Physics, Budapest HU-1117}

\corres{Correspondence: szigeti.balazs@wigner.hu}

\firstnote{barna.imre@wigner.hu} 
\secondnote{barnafoldi.gergely@wigner.hu}

\abstract{We present a dark fluid model described as a non-viscous, non-relativistic, rotating, and self-gravitating fluid. We assumed that the system has spherical symmetry and the matter can be described with the polytropic equation of state. The induced coupled non-linear partial differential equation system was solved by using a self-similar time-dependent {\it ansatz} introduced by L. Sedov and G. I. Taylor. These kinds of solutions were successfully used to describe blast waves induced by an explosion since the Guderley–Landau–Stanyukovich problem. We showed that the result of our quasi-analytic solutions are fully consistent with the Newtonian cosmological framework. We analyzed relevant quantities from the model: the evolution of the Hubble parameter and the density parameter ratio. We have found that our solutions can be applied to describe normal-to-dark energy on the cosmological scale.}


\begin{document}


\section{Introduction}

In the second half of the 20\textsuperscript{th} century, various self-similar solutions have been found after Gottfried Gurderley's famous discovery of spherically symmetric self-similar solutions that describe an imploding gas that collapses to the center~\cite{ref-Guderley}. In this paper, we used those kinds of self-similar solutions which were found by Leonid Ivanovich Sedov and Sir Geoffrey Ingram Taylor independently during the 1940s~\cite{ref-Sedov,ref-Taylor}. Despite the fact that such models are well-known for decades they have recently received attention again. This \emph{ansatz} has been already applied successfully in several hydrodynamical systems, like the 3-dimensional Navier-Stokes and Euler equations~\cite{ref-Barna1}, and heat equation~\cite{ref-Bressan,ref-Barna2}, or star formation~\cite{ref-Guo}. Also, the concept of self-similarity has a wide range of applications in general relativity. Homothetic solutions were first introduced by Cahill and Taub \cite{ref-cahill}, and have been studied extensively for different topics, such as asymptotic solutions in cosmology \cite{ref-cosmology}, and the gravitational collapse of black holes \cite{ref-collapse}.

The existence of the dark matter was first proposed by the Dutch astronomer Jacobus Cornelius Kapteyn~\cite{ref-Kapteyn} and became widely known through Zwicky's famous work from 1933~\cite{ref-Zwicky}. During the second half of the century, solid experimental evidence was provided by Vera Rubin, Ken Ford, and others~\cite{ref-Rubin,ref-Einasto}. However, the general existence and specified properties of dark matter, are still one of the most disputed topics in theoretical astrophysics. Dark fluid is one of the theoretical attempts to describe the properties of dark matter and its unification with dark energy into one hypothesized substance~\cite{ref-Arbey}. 

Our goal is to use the Sedov-Taylor \emph{ansatz} to describe the time evolution of a dark fluid-like material characterized by a coupled, non-linear partial differential equation system. In our model, we studied one of the simplest dark fluid material described by a polytropic (linear) equation of state. The dynamical evolution of the dark fluid is governed by the Euler equation and the gravitational field is described by the corresponding Poisson equation. We found time-dependent scaling solutions of the velocity flow, density flow, and gravitational fields, which can be good candidates to describe the evolution of the gravitationally coupled dust-like dark matter in the Universe. We showed that these kinds of solutions are consistent with the Newtonian Friedmann equations. They satisfy the mass conservation and acceleration equation and provide a similar expansion rate of the Universe as the traditional solutions.  The aim of this study was to broaden our knowledge of time-dependent self-similar solutions in these dark fluid models, which improve and extend our previous model~\cite{ref-Pocsai}. We tested our model on the cosmological scales. Moreover, we studied how the obtained evolution of the Hubble parameter and the ratio of the density parameters resemble other physical models.
\section{The Model}

We consider a set of coupled non-linear partial differential equations, which describes the non-relativistic dynamics of a compressible fluid with zero thermal conductivity and zero viscosity~\cite{ref-Euler}, 
\begin{subequations}
\begin{align}
    \partial_t \rho + \mathrm{div} (\rho \boldsymbol{u}) &= 0 \label{eq:01A} \ , \\
    \partial_t (\rho \boldsymbol{u}) + \mathrm{div} ( \rho \boldsymbol{u} \otimes \boldsymbol{u} )  &= - \nabla P(\rho) + \rho \boldsymbol{g} \ , \label{eq:01B}\\
    P &= P (\rho) \ . \label{eq:01C}
\end{align}
\label{eq:01}
\end{subequations}
These equations are the continuity, the Euler equation, and the equation of state (EoS), respectively. We assume that the system has spherical symmetry and we are interested in solving it in one dimension. If we imply that the fluid is ideal and the system has spherical symmetry we can reduce the multi-dimensional partial differential equation (PDE) system into the one-dimensional, radial-dependent one  
\begin{subequations}
\begin{align}
    \partial_t \rho + (\partial_r \rho) u + (\partial_r u) \rho + \dfrac{2u\rho}{r} &= 0 \label{eq:02A} \ , \\
    \partial_t u + (u \partial_r) u &= - \dfrac{1}{\rho} \partial_r P + g \label{eq:02B} \ , \\
    P& = P(\rho) \label{eq:02C} \ .
\end{align}
\label{eq:02}
\end{subequations}
Here, the dynamical variables are the $\rho = \rho(r, t)$, $u = u(r, t)$, and $P = P(r, t)$ which means the density, the radial velocity flow, and the pressure field distributions, respectively. The $g$ is the radial component of an exterior force density. As we presented briefly in the introduction we used the following general linear equation of state
\begin{equation}
    P(\rho) = w \rho^n, \quad \quad n = 1 \ .
\label{eq:03}
\end{equation}

Several forms of the EOS are available in astrophysics and polytropic ones were successfully used in the past, see Emden's famous book~\cite{ref-Emden}. A great variety of applications can be found in Ref.~\cite{ref-Horedt}. In the equation Eq.~\eqref{eq:03}, the $w$ parameter can vary depending on the type of matter that governs the system's evolution. Traditionally, the $w=0$ is used which value corresponds to the EoS for ordinary non-relativistic matter or cold dust. For our case, we can also choose a negative value for the $w$ which leads us to different kinds of dark-fluid scenarios as was presented in detail by Perkovic~\cite{ref-Perkov}. In this paper, we chose $w=-1$ which represents the simplest case of the expanding universe governed by dark matter. Smaller values could cause the Big Rip. The adiabatic speed of sound can be evaluated from Eq.~\eqref{eq:03} and it is easy to show that it will be constant.
\begin{equation}
    \dfrac{\mathrm{d}P(\rho)}{\mathrm{d}\rho} = c_s^2 = w  \ , 
\end{equation}
%
which is a necessary physical condition. Furthermore, let us assume that we have an additional self-gravitating term in the Eq.~\eqref{eq:02B}. In this case, the exterior force density, $g$ can be expressed in the following way: 
\begin{equation}
    \boldsymbol{g} = - \partial_r \Phi \ , 
\label{eq:05}
\end{equation}
%
where the $\Phi=\Phi(r,t)$ is the Newtonian gravitational potential and it satisfies the Poisson equation which will couple to the previously proposed PDE system~\cite{ref-Poisson},   
\begin{equation}
    \nabla^2 \Phi = 4 \pi G \rho  \ , 
\end{equation}
where the $G$ is the universal gravitational constant which is set to unity in further calculations. One can notice that we can also add an additional constant term $\Lambda$ to the Eq.~\eqref{eq:01B} which has a similar role as the cosmological constant in Einstein's equations 
\begin{equation}
    \partial_t u + (u \partial_r) u = - \dfrac{1}{\rho} \partial_r p - \partial_r \Phi(r) + \Lambda \ . 
\label{eq:06}
\end{equation}

We are going to show below that, this constant cannot be used since it does not lead to a consistent self-similar solution, which is what we are looking for. Note, this observation in our model can be an {\it indirect proof of the nonexistence of the static Universe picture}. We can extend the exterior force density further with a rotating term. In this case, we would like to add a phenomenological rotation term to the Eq.~\eqref{eq:02B}, thus the equation will take the following modified form
\begin{equation}
    \partial_t u + (u\partial_r)u = - w\dfrac{1}{\rho} \partial_r \rho  - \partial_r \Phi(r) + \dfrac{\sin \theta \omega^2r}{t^2} \ , 
\end{equation}
%
where $\omega$ is a dimensionless parameter that describes the strength of the rotational effect and $\theta$ is the polar angle. We assumed that the rotation is slow, therefore we can expect that the spherical symmetry is not broken. This statement is satisfied if the $\omega$ parameter is sufficiently small, implying that the rotational energy is negligible compared to the gravitational energy.  The self-similar analysis of various rotating and stratified incompressible ideal fluids were investigated in two Cartesian coordinates~\cite{rotstrat}. Note that for the calculations below, the geometrized unit system ($c = 1$, $G=1$) was applied, which can be converted to other units. See Appendix A for more details and unit conversations. 

\section{Scaling Solution and Sedov-Taylor \emph{Ansatz}} \label{sec::Sedov}

We would like to find and study analytic solutions of the equations by applying the long-established self-similar \emph{ansatz} by Sedov and Taylor~\cite{ref-Sedov,ref-Taylor} which can be expressed in the following form  
\begin{subequations}
\begin{gather} 
    u (r,t)  = t^{-\alpha} f \bigg( \dfrac{r}{t^{\beta}} \bigg) \ ,
    \label{eq::SedovTaylorAnsatzA}\\
    \rho(r,t) = t^{-\gamma} g \bigg( \dfrac{r}{t^{\beta}} \bigg) \ , \label{eq::SedovTaylorAnsatzB}\\
    \Phi(r,t) = t^{-\delta} h\bigg( \dfrac{r}{t^{\beta}} \bigg), \label{eq::SedovTaylorAnsatzC}
\end{gather}
\label{eq::SedovTaylorAnsatz}
\end{subequations}
where the $r$ means radial and $t$ means time dependence. One can notice that the so-called shape functions $(f,g,h)$ only depend on the $rt^{-\beta}$, thus we introduce a new variable
\begin{equation}\label{eq::Zeta}
\zeta = r t^{-\beta}.  
\end{equation}
The $\zeta$ is a dimensionless quantity in geometrized unit. The not yet determined exponents are called similarity exponents ($\alpha$, $\beta$, $\gamma$, and $\delta$) and they indeed have physical relevance. The $\beta$ describes the rate of spread of the spatial distribution during the time evolution if the exponent is positive or the contraction if $\beta < 0$. In addition, the other exponents describe the rate of decay of the intensity of the corresponding field. Solutions with integer exponents are called self-similar solutions of the first kind, while the second kind denotes the non-integer ones. Self-similarity is based on the concept that physical quantities will preserve their shape during time evolution. A general description of the properties of these types of scaling solutions can be found in our previous publication~\cite{ref-Pocsai}.

We assumed that the shape functions are sufficiently smooth and it is at least continuously differentiable (twice) in $\zeta$ over the entire domain. Thus, we have calculated the relevant time and space derivatives of the shape functions and substituted them into the equations~\eqref{eq:02}. As a consequence, we got usually an overdetermined algebraic equation system for the similarity exponents. Other possible scenarios may play out and these were presented in detail in this paper~\cite{ref-Pocsai}. We obtained the following numerical value for the exponents $\alpha = 0$, $\beta = 1$, $\gamma = 2$, and $\delta = 0$ for both the non-rotating and the rotating cases. If we add the $\Lambda$ constant to the Euler equation, such a solution for the similarity equation cannot be found. From the results, it is evident that the dynamical variables such as the velocity, gravitational potential, and density flow have spreading properties. Our physical intuition says that spreading is somehow similar to expansion which is a basic property in the Universe at astronomical, or cosmological scales.

\begin{figure}[H]
\centering
\includegraphics[width=0.6\textwidth]{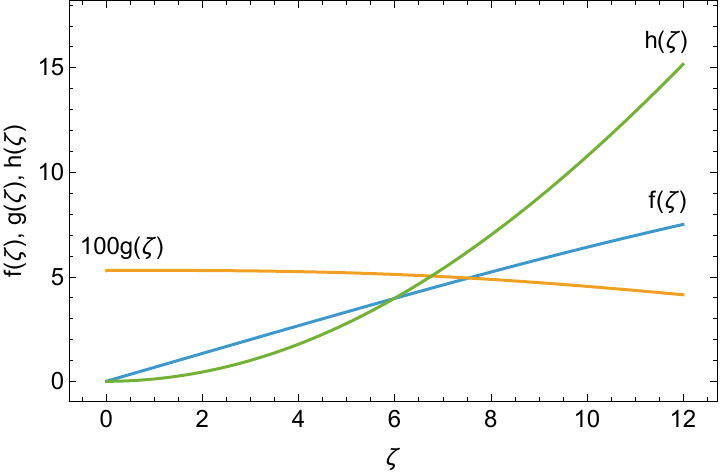}
\caption{Numerical solutions of the shape functions, the integration was started at $\zeta_{0} = 0.001$, and the initial conditions of $f(\zeta_{0}) = 0.05$, $g(\zeta_{0}) = 0.053$, $h(\zeta_{0}) = 0$, and $h'(\zeta_{0}) = 1$ were used. For better visibility, the function $g(\zeta)$ was scaled up with a factor of 100. The values are given in geometrized unit. 
\label{fig1}}
\end{figure}

By substituting the obtained numerical values of the similarity exponents, we have reduced the induced PDE system into an ordinary differential equation (ODE) system that depends only on the $\zeta$ independent variable. We found that the obtained equation system has the following form,
\begin{subequations}
\begin{align}
    - \zeta g'(\zeta) + [f'(\zeta) - 2] g(\zeta) + f(\zeta) g'(\zeta) + \dfrac{2f(\zeta)g(\zeta)}{\zeta} &= 0, \label{eq::11A}\\ 
    -\zeta^2 f'(\zeta) + \zeta f'(\zeta) f(\zeta) & = - \dfrac{w g'(\zeta)}{g(\zeta)}  \zeta - h'(\zeta) \zeta + \omega^2\sin\theta \zeta^2,\label{eq::11B}\\
     2h'(\zeta) + h''(\zeta) \zeta & = g (\zeta) 4 \pi G \zeta \ . \label{eq::11C}
\end{align}
\label{eq::11}
\end{subequations}
One can easily notice that the presented ordinary differential equation system Eq.~\eqref{eq::11} cannot be solved analytically. For linearized non-autonomous ordinary differential equation systems, the stationer point of the phase space can be found as well as one can say something about the general asymptotic behavior of the solutions~\cite{ref-Kenneth}. Nonetheless, there is no generally known method for non-linearized non-autonomous differential equation systems. Also, the existence and uniqueness of smooth solutions have not yet been proven in multiple dimensions. Therefore, it is a reasonable approach to solve the obtained ordinary differential equation system, Eqs.~\eqref{eq::11} numerically for a large number of parameter sets (based on physical considerations) to explore the behavior of the solution of the system with different boundary- and initial conditions. One example of the numerical solution can be seen in Fig.~\ref{fig1}. 

As an example, at a specific parameter and initial condition set, the shape function of the velocity, $f(\zeta)$ is almost linear after a short decrease and increasing giving a hint for a Hubble expansion-like behavior. The shape function, $g(\zeta)$ is asymptotically flat after a quick ramp-up, corresponding to the conservation of matter. The last shape function, $h(\zeta)$ has an increasing polynomial trend with a slight positive exponent, connected to the gravitational potential. To obtain a sufficiently smooth numerical solution we solved the ODE system by using an adaptive numerical integration provided by \emph{Wolfram Mathematica 13.1} \cite{ref-Wolfram}. For all of our calculations, the integration limits were $\zeta_{0} = 0.001$ and $\zeta_{max} = 40$ as in Ref.~\cite{ref-Pocsai}. As was said before, we established some initial conditions to obtain the numerical solution, due to this reason we have used ranges, $\mathcal{R}$ of $f(\zeta_0) = 0.005-0.5$, $g(\zeta_0) = 0.001-0.1$ and for the second order differential equation, we have the $h(\zeta_0) = 0$ and $h'(\zeta_0) = 1$.
This choice of initial condition reflects that firstly it is physically reasonable that the density flow range is $\mathcal{R}(g) \subset \mathbb{R}^+$ and finite. Some recent results suggest that dark fluid can possibly have negative mass~\cite{ref-Farnes}, but in this model, this leads us to singular solutions. Secondly, this choice of the initial velocity flow $\mathcal{R}(f) \subset \mathbb{R}^+$ means an initially radially expanding fluid. We have seen that if the initial value for $f(\zeta)$ and $g(\zeta)$ were set outside of the previously given range, the solution of the differential equation becomes singular. We also saw that the variation in the initial condition corresponding to the shape function of the gravitational potential does not affect the trend of the time-evolution of the system it only causes vertical shifts. Therefore, we set the initial numerical value equal to zero.

We are interested to find the solution for the ODE system as a function of the spatial and time coordinates. We transform our one-variable numerical solutions into two-variable functions, for this we used the inverted form of the Eq.~\eqref{eq::Zeta}. One can easily notice if we look at the shape of the \emph{ansatz} that the solution will have a singularity at $t=0$. Thus, we used the $0.001 \leq t \leq 25$ and $0.001 \leq r \leq 25$ domains to obtain the space and time-dependent initial dynamical functions, $u(r,t)$, $\rho(r,t)$, and $\Phi (r,t)$.

\section{Results}
\label{sec:res}

Here, we present the solutions of the self-gravitating non-relativistic dark fluid. Firstly, we give a detailed introduction to the global properties of the solutions in a non-rotating system. Secondly, we will show the effect of the slow rotation on the solutions. In addition to that we compare the results from the two cases with each other and with the previous results in Ref.~\cite{ref-Pocsai}. Note, the spherical symmetry of the system was kept conserved for all the cases.

\subsection{Non-rotating system}
\begin{figure}[H]
\includegraphics[width=\textwidth]{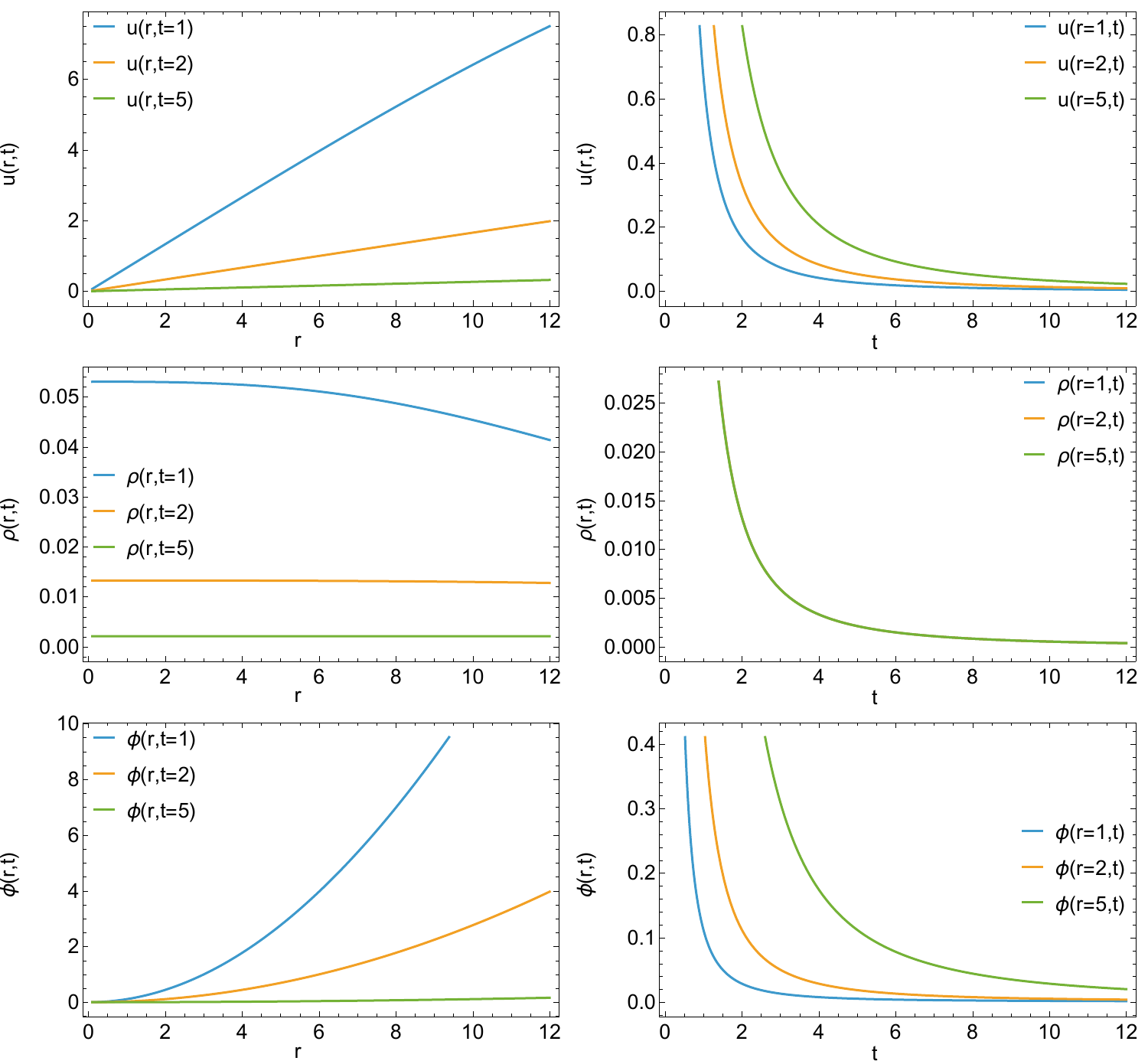}
\caption{ Different radial (left) and time (right) projections of the velocity flow (1\textsuperscript{st} row), density (2\textsuperscript{nd} row), and gravitational potential (3\textsuperscript{rd} row) for the non-rotating case, respectively. A detailed explanation is given in the main text. The domain range is given in geometrized unit.\label{fig2}}
\end{figure}

\begin{figure}[H]
\centering
\includegraphics[width=\textwidth]{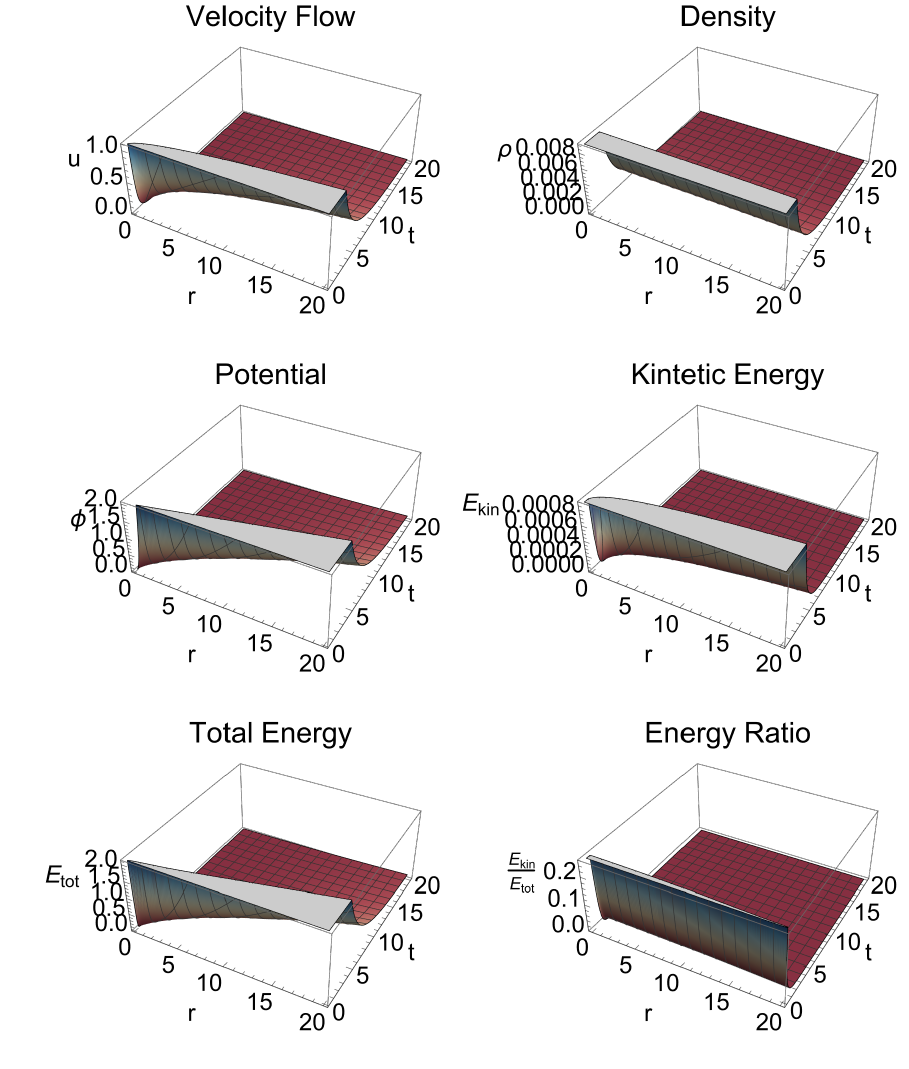}
\caption{Numerical solutions of the velocity flow $u(r,t)$, density flow $\rho(r,t)$, and gravitational potential $\Phi(r,t)$ as a function of the spatial and time coordinates in case of a non-rotating system. We also present the distribution of the total and kinetic energy density. For the numerical integration, we used $\zeta_{0} = 0.001$, and the initial conditions were $f(\zeta_{0}) = 0.05$, $g(\zeta_{0}) = 0.053$, $h(\zeta_{0}) = 0$, and $h'(\zeta_{0}) = 1$.  \label{fig3}}
\end{figure}

In the first case, we set the $\omega$ parameter to zero and we used the obtained numerical values for the similarity exponents ($\alpha$, $\beta$, $\gamma$, and $\delta$) to obtain the exact ordinary differential equation. For the numerical integration, we used the initial condition  $f(\zeta_0) = 0.05$ and  $g(\zeta_0) = 0.053$ for the velocity and density flow respectively. Firstly, we used time and radial projection of the unknown functions for better understanding. Fig.~\ref{fig2} illustrates the spatial and time projection of the obtained velocity, density, and gravitational potential. These velocity flow and density flow results are consistent with our initial statement that these kinds of solutions of dark fluid can be used as a model to describe the exploding system (e.g. the Universe). We can see similar behavior for the radial velocity and the density, they have a quick decay in time at all distances. Also, they have a real singularity at $t = 0$, due to the shape of the {\it ansatz}. However, the radial distribution shows different nature. The density is constant near the center of the explosion distances and becomes linear at large distances. On the contrary, the velocity grew polynomially with the radial distance. One can see, that the gravitational potential decreases hyperbolically over time, and becomes asymptotically flat. The gravitational potential has a natural singularity at the origin. The local existence of global weak solutions on a domain outside of the origin has been shown by Tsunge and others, despite the existence of a singularity. 
Compared to the previous, non-rotating, model with two-equations presented in Ref.~\cite{ref-Pocsai}, we found a different radial velocity profile. Also, the similarity exponents are different $\alpha = 0$, $\beta = 1$, and $\gamma = -1$ in the two-equation model. This is most likely due to the new smoother solution forming as a consequence of the effect of the second derivative appearing in the Poisson equation. Furthermore, we have seen that the solution depicted above in Fig.~\ref{fig2} is numerically stable in the specified initial and boundary condition range. It is more relevant to investigate the dynamics of the complete fluid in time and space to understand some general trends or physical phenomena as the function of the initial conditions. Due to this reason, we evaluated the related energy densities, which are the following
\begin{equation}
    \epsilon_{kin} (r,t) = \dfrac{1}{2} \rho (r,t) u^2(r,t) , \qquad \Phi (r,t) = h(r,t), \qquad \epsilon_{tot}(r,t) = \epsilon_{kin}(r,t) + \Phi(r,t).
\end{equation}
Fig.~\ref{fig3} illustrates that the kinetic energy density has a singularity at $t=0$ as we have seen in the case of radial velocity. It has linearly enhancing maxima at larger distances and has a quick decay in time for all radial distances. As we mentioned above, the gravitational potential is a negative polynomial in time. Thereby, we can obtain the total energy density of the system. From the total energy density distribution, it is apparent that the short-time behavior of the system is predominated by the initial explosion and the long-range structure is regulated by the gravitational potential. 
\subsection{Rotating system}
\begin{figure}[H]
\centering
\includegraphics[width=\textwidth]{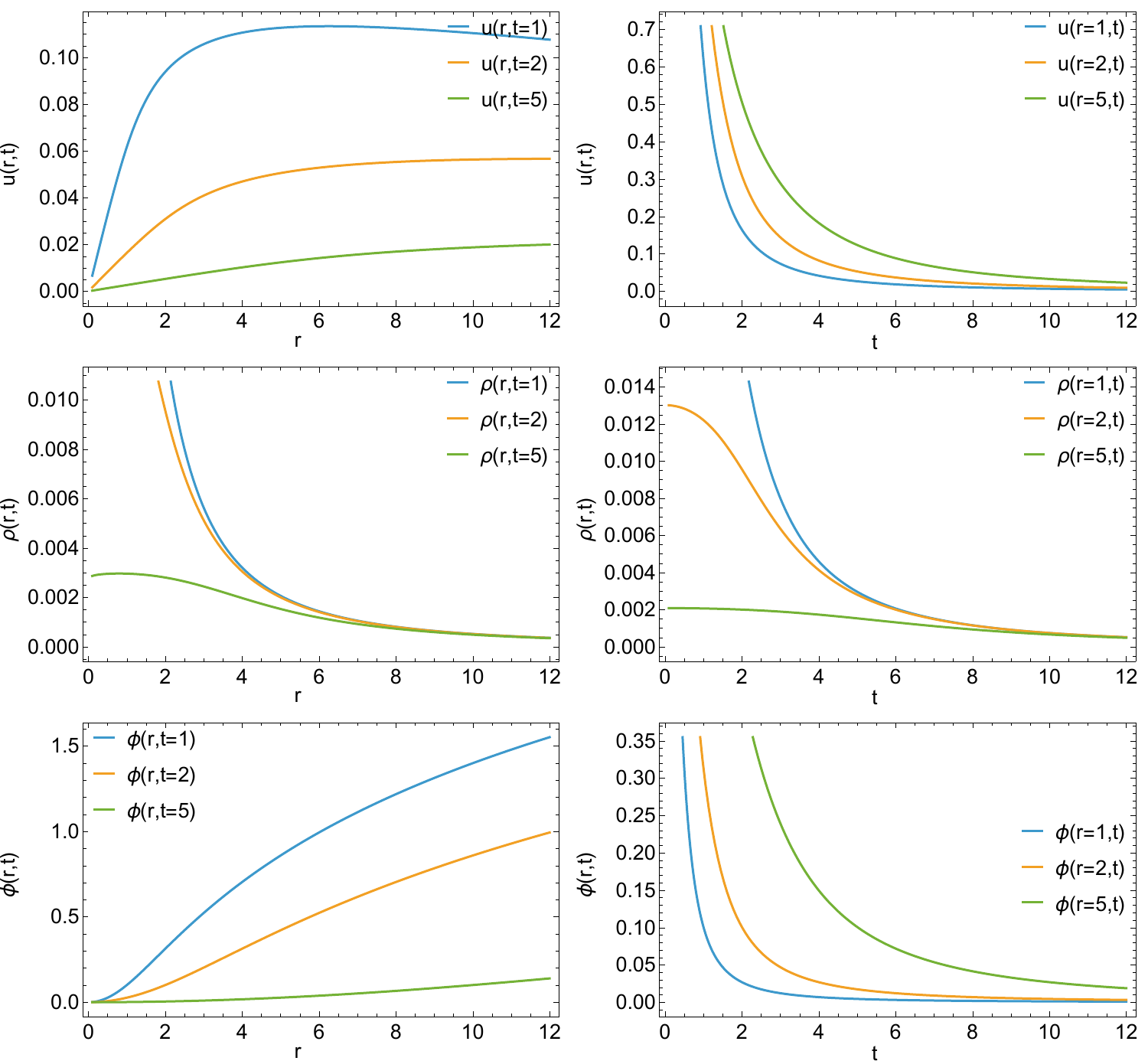}
\caption{The time and radial projections of the velocity flow (1\textsuperscript{st} row), density (2\textsuperscript{nd} row), and gravitational potential (3\textsuperscript{rd} row) respectively for the rotating system $(\omega = 0.1535)$. For the numerical integration we used $\zeta_{0} = 0.001$, and the initial conditions were $f(\zeta_{0}) = 0.05$, $g(\zeta_{0}) = 0.053$, $h(\zeta_{0}) = 0$, and $h'(\zeta_{0}) = 1$. \label{fig4}}
\end{figure} 
In this section, we analyzed the effect of slow rotation compared to the non-rotating case. Firstly, we studied the effect of the variation of the maximal angular velocity, $\omega$ parameter. We have chosen the  polar angle, $\theta$ parameter at the equatorial, which gives the largest effect. As previously stated we assumed for further analysis that the spherical symmetry is not broken. According to that, we fixed that the gravitational force density is at least a magnitude larger than the centrifugal force density at every time and space $(\|f_{grav}\| \gg \|f_{centr}\|)$. Numerical results showed us that an $\omega$ range can be found where the constraint will be fulfilled if the previously specified initial condition set is valid. We demonstrate that the asymptotic behavior of the numerical solution has a significant $\omega$ dependence on the acceptable (0 < $\omega$ < 0.3) domain of parameters and initial conditions.

If we compare the results shown in Fig.~\ref{fig4} with the non-rotating case (Fig.~\ref{fig2}), it is evident that slow and constant rotation does not affect the time and spatial distribution of the gravitational potential. Moreover, we can see that the radial density profile of the system is nearly uniform  at large distances and identically to the previous case it decreases rapidly over time.  Thus we can conclude, that the rotation accelerates the even distribution of the material in space and speeds up inflation. 
\newpage
The singular behavior close to the $t = 0$ does not affect by the rotation as was expected. However, a significant difference can be seen as one looks at the first graph (top  left panel of Fig.~\ref{fig4}). One can see that the radial profile of the velocity flow starts from zero in the origin and it shows exponential growth for the short-range behavior.
\begin{figure}[H]
\centering
\includegraphics[width=\textwidth]{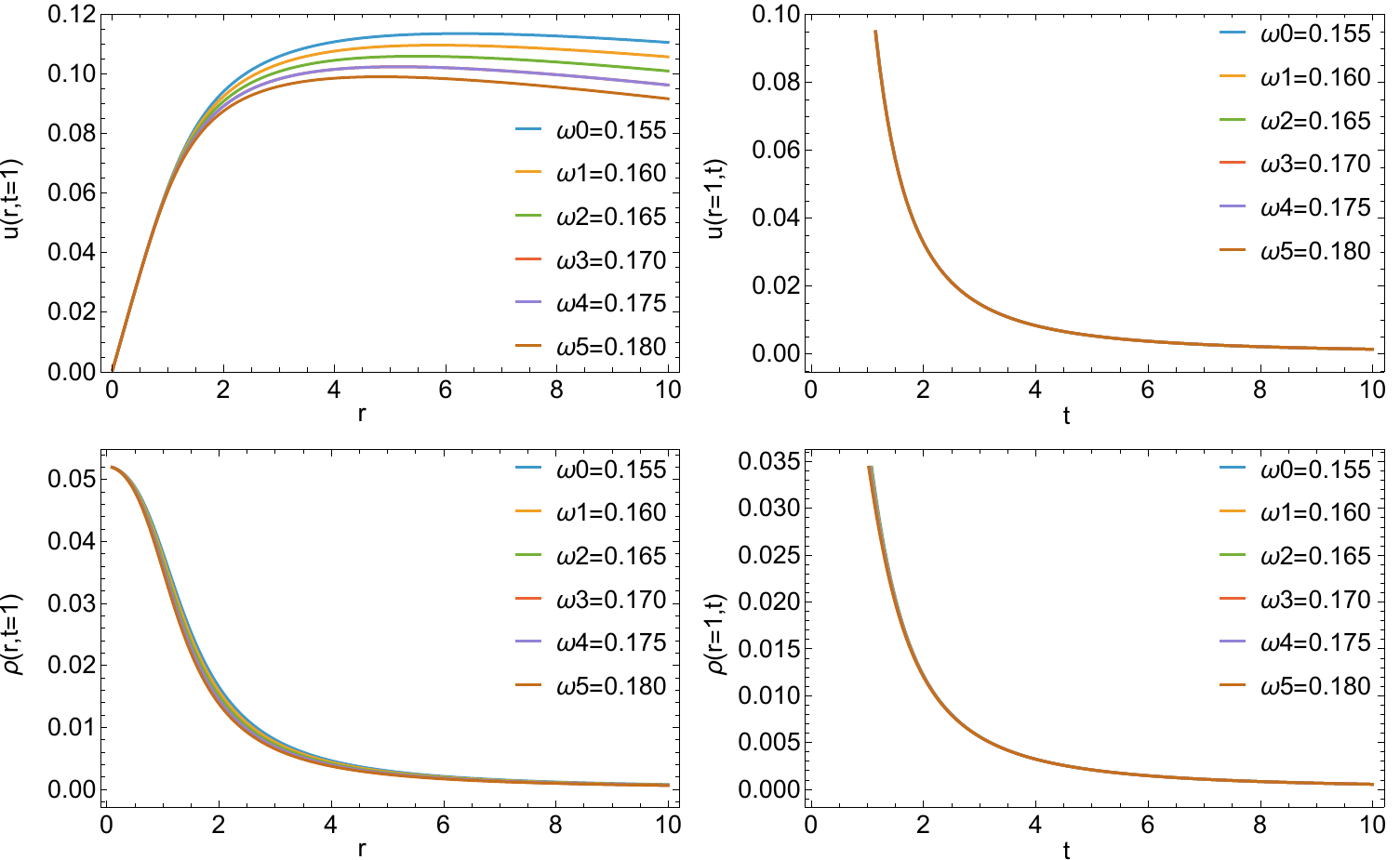}
\caption{The maximal angular velocity $\omega$ dependence of the space and time evolution. Different lines correspond to different angular velocity values, $\omega$. The curves were evaluated at a particular time (left) and radial (right) coordinates were given on the vertical axis. A detailed explanation is given in the main text. \label{fig5}}
\end{figure}  
Fig.~\ref{fig5} illustrates that the $\omega$ value has a critical influence on the long-range asymptotic behavior of the time evolution of the velocity flow. Moreover, the increase of the $\omega$ value causes significant modification in the radial profile for both the velocity and density flow but leaves the time evolution unaltered. In the analysis of the behavior of the obtained numerical solutions, we found that on the inspected initial value range shows similar behavior. An $\omega < 1$ can be found for every initial and boundary condition where the long-range asymptotic structure alternates, an example of this can be seen in Fig.~\ref{fig5}. Likewise, in the previous case, we studied the properties of the relevant dynamic variables (Fig.~\ref{fig6}). The energy density associated with rotation and the total energy is
\begin{equation}
    \epsilon_{rot}(r,t) = \dfrac{1}{2} \rho (r,t) \omega^2 r \quad\quad \epsilon_{tot}(r,t) = \Phi(r,t) + \epsilon_{kin}(r,t) + \epsilon_{rot}(r,t).
\end{equation}
\begin{figure}[H]
\centering
\includegraphics[width=0.85\textwidth]{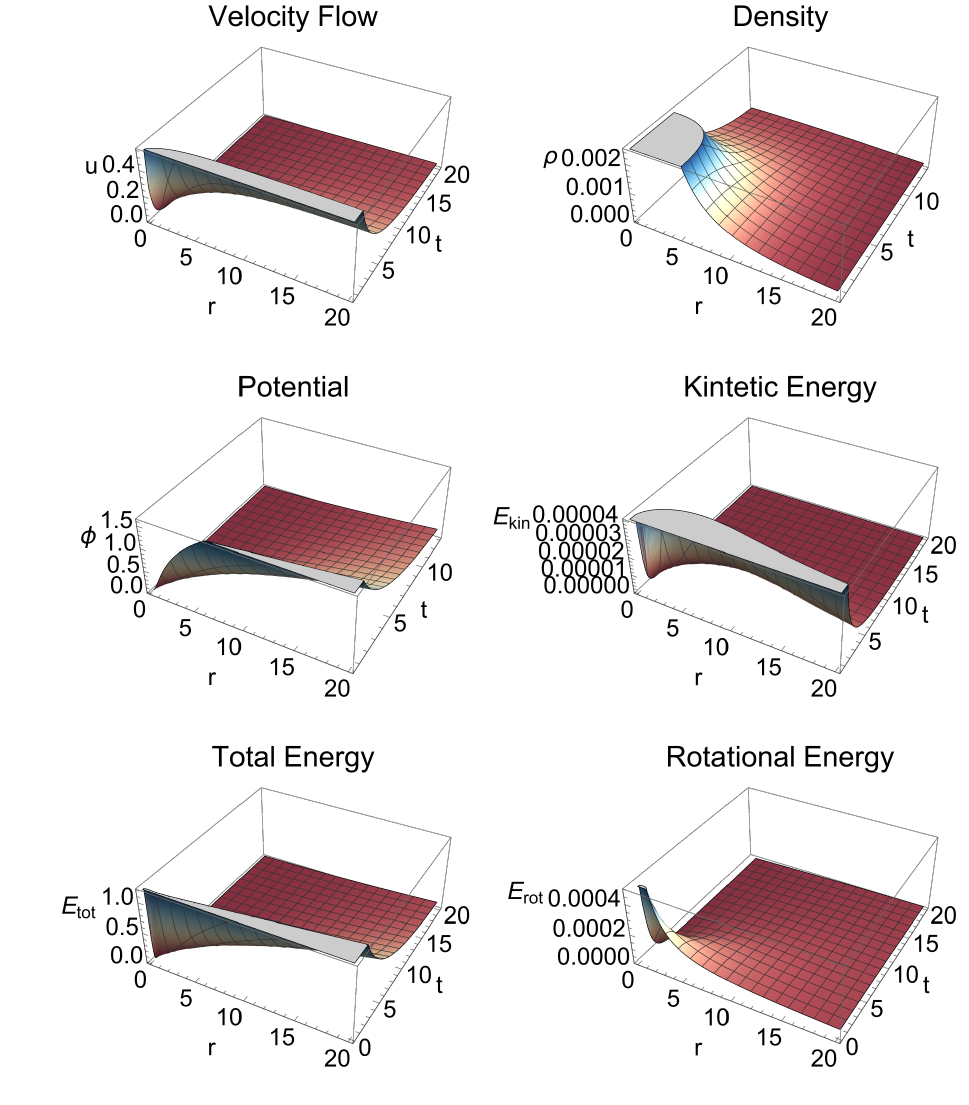}
\caption{Numerical solutions of the velocity flow $u(r,t)$, density flow $\rho(r,t)$, and gravitational potential $\Phi(r,t)$ as a function of the spatial and time coordinates for the rotating case. We also present the distribution of the total and kinetic energy density. For the numerical integration we used $\zeta_{0} = 0.001$, and the initial conditions were $f(\zeta_{0}) = 0.05$, $g(\zeta_{0}) = 0.0{\color{red}53}$, $h(\zeta_{0}) = 0$, and $h'(\zeta_{0}) = 1$. We have used the $\omega = 0.1535$ parameter. \label{fig6}}
\end{figure} 

\section{Connection to Newtonian Friedman Equation}

We apply the obtained scaling solution to cosmology and explain the evolution of the Universe. Our strategy is to describe a concept of cosmic fluid in Newtonian cosmology, which is applied since relativistic effects are less significant.  In this case, there is no need for a global reference point. Thus a \emph{scale-factor} $a(t)$ can be induced which contains all of the time evolution that affects our chosen reference frame. Henceforward, the relative distances 
in time can be denoted in this term,
\begin{equation}
R(t) = a(t) \, l,   
\end{equation}
where $R(t)$ is a continuous function  and $l$ is real number. In this chapter, we have used classical conservation equations to show the connection between the obtained self-similar solution and the traditional Friedmann equations.

\subsection{Connection to the expansion rate} \label{sec::ExpansionRate}

We introduce the total mass inside a radius of $r$,
\begin{equation} \label{eq::masscons}
   M(t) = \int_{{\cal V}(t)} \rho(R(t),t) \, \mathrm{d}V = 4\pi \int_0^r \rho(R(t),t) \, R(t)^2 \,\mathrm{d}R(t),  
\end{equation}
where ${\cal V}(t) \subset \mathbb{R}^3$ is a sphere with radius $R(t)$ and $ r \in (0,R(t))$. Within the classical Newtonian framework, therefore the energy and mass conservation principle are fulfilled separately. 
\begin{equation}
    \dfrac{\mathrm{d}}{\mathrm{d}t}M(t) = 4 \pi  \dfrac{\mathrm{d}}{\mathrm{d}t} \int \rho(a(t)l,t)  \, a^3(t) \ l^2  \, \mathrm{d}l \overset{!}{=} 0. 
\end{equation}
Replacing the derivation with the integral formula one can assume that equality holds for every $l$ 
\begin{equation}
    \rho(a(t)l,t){\color{red}^{-1}} \dfrac{\mathrm{d}}{\mathrm{d}t} \bigg[\rho(a(t)l, t)\bigg] = - 3\dfrac{\dot{a}(t)}{a(t)}.
\end{equation}
We also impose the kinematic condition that,
\begin{equation} \label{eq::FriedmannSelfSimilar}
    \dfrac{\mathrm{d}}{\mathrm{d}t}R(t) = u(R(t),t) \Rightarrow \dfrac{1}{g(R(t),t)} \dfrac{\mathrm{d}}{\mathrm{dt}} \bigg[ t^{-\gamma} g(R(t),t)\bigg]  = -3\dfrac{t^{-\alpha}f(R(t),t)}{R(t)}. 
\end{equation}
We impose that $\rho(r,t)>0 \quad $ for $ \forall r \geq R(t)$ and we assumed that the velocity $u(r,t)$ and density $\rho(r,t)$ is obtained from the self-similar {\it ansatz}, Eq.~\eqref{eq::SedovTaylorAnsatz}. We expand the density and velocity flow into power series:  
\begin{equation}
    \rho(r,t) \sim t^{-\gamma} \sum_{m=0}^{\infty} \rho_{m} \zeta^{m} \quad \text{and} \quad u(r,t) \sim t^{-\alpha} \sum_{n=0}^{\infty} u_n \zeta^n.
\end{equation}
After substituting $\zeta = R(t) t^{-\beta}$, the Eq.~\eqref{eq::FriedmannSelfSimilar} becomes,
\begin{equation} \label{eq::Assumption}
\begin{split}
        - \gamma t^{-(\gamma + 1)} + t^{-\gamma} \sum_{m=1}^{\infty} \rho_{m} m \big(R(t)t^{-\beta}\big)^{m-1} \big(\Dot{R}(t)t^{-\beta} - \beta R(t) t^{-(\beta+1)} \big) = \hspace{2truecm} \\ = - \dfrac{3}{R(t)} \bigg[ t^{-\alpha} \sum_{n=0}^{\infty} u_n \big(R(t)t^{-\beta}\big)^n  \bigg] \bigg[ t^{-\gamma} \sum_{m=0}^{\infty} \rho_m \big(R(t)t^{-\beta}\big)^m  \bigg].  
\end{split}
\end{equation}
On the relevant space and time scale, we only considered the contributions of the finite term. The discussion of the rotating case in Section~\ref{sec:res} requires that we have to use a high-order polynomial to approximate the $f(\eta)$ function. Analysing the numerical results on Fig.~\ref{fig4}, we can restrict ourself to use polynomes up to the 8\textsuperscript{th} order, which can approximate  curves well.
\begin{equation}
   u(r,t) \sim t^{-\alpha} \sum_{n=0}^8 \Tilde{u}_n \zeta^n \quad \text{and} \quad \rho(r,t) \sim t^{-\gamma} \sum_{m=0}^8 \Tilde{\rho}_m \zeta^m  \ .
\end{equation}
Moreover, within the domain of interest a further simplification can be applied for $\rho$ describing it as a polynomial with rational exponent $\rho(\zeta) \sim A \zeta^{\kappa}$, where $\kappa = 6/7$. Hence the Eq.~\eqref{eq::Assumption} can be rewritten, 
\begin{equation}
    \kappa \Dot{R}(t) - \dfrac{1}{t}[\gamma + \kappa \beta] R(t) + 3 t^{-\alpha} \sum_{n=0}^8 \Tilde{u}_k ( R(t) t^{-\beta} )^n = 0.
\end{equation}
Consequently, this mass conservation equation~\eqref{eq::masscons} becomes a non-autonomous first-order differential equation. This cannot be solved analytically due to the high non-linear terms. 

\begin{figure}[H]
\centering
\includegraphics[width=0.6\textwidth]{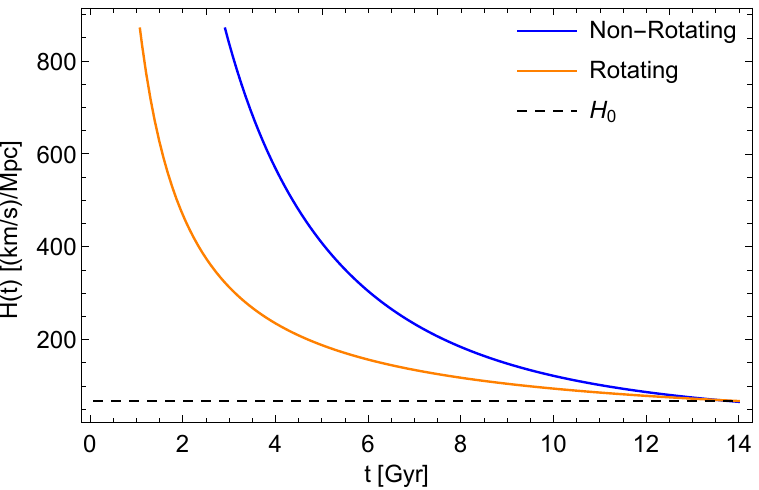}
\caption{Analytical (non-rotating) and numerical (rotating) solutions of the expansion rate of the Universe, the integration was started at $\zeta_{0} = 0.001$, and the initial conditions of $f(\zeta_{0}) = 0.05$, $g(\zeta_{0}) = 0.01$, $h(\zeta_{0}) = 0$, and $h'(\zeta_{0}) = 1$ were used. The results match well with the data from the~\cite{ref2-Hubble}. 
\label{fig::HubbleTimeEvolution}}
\end{figure}

In the non-rotating limit ($\omega \to 0$), further simplifications can be done, since the higher-order terms will be less and less significant, therefore the velocity flow becomes simply $u(r,t) \sim u_1 \zeta^1 + u_2 \zeta^2$ and the mass conservation equation is
\begin{equation} \label{eq::DiffScaleFactor}
     \kappa \Dot{R}(t) + 3u_2t^{-(\alpha + 2 \beta)}[R(t)]^{2} - \dfrac{1}{t}[\gamma + \kappa \beta] R(t) + 3 u_1 R(t) t^{-(\alpha+\beta)} = 0.
\end{equation}
It can be solved analytically and the solution is
\begin{equation} \label{eq::GeneralSolution}
\begin{split}
       R(t) = u_1  t^{\beta +\frac{\gamma }{\kappa }} \exp\left[ -\frac{3 u_1 t^{\mu}}{\mu \kappa }\right] \times \bigg[ 3^{-\frac{\gamma }{\mu \kappa }} u_2 t^{\gamma /\kappa } \left(\frac{u_1 t^{\mu }}{\nu }\right)^{-\frac{\gamma }{\mu \kappa }} \Gamma \left(1+\frac{\gamma}{\nu },\frac{3 u_1 t^{\mu}}{\nu }\right)- \mathcal{H}_1 u_1 \bigg]^{-1}, 
\end{split}
\end{equation}
where $\mu = 1 - (\alpha + \beta)$ and $\nu = \kappa - \beta \kappa$, the $\alpha,\beta,\gamma$ are the similarity exponents and $A,u_1,u_2$ positive constants and $\kappa$ positive exponent is obtained from the solution of the Eq.~\eqref{eq::11}. The $\mathcal{H}_1$ is an integration constant and $\Gamma()$ is the upper incomplete Gamma function \cite{NIST}. The Eq.~\eqref{eq::GeneralSolution} can be simplified if we substitute the similarity exponents and constants from the non-rotating case,
\begin{equation}
    R(t) = \frac{t}{\mathcal{H}_1 t^{\frac{3 u_1-2}{\kappa }}+\frac{3 u_2}{2-3 u_1}}, \quad \text{where } \kappa = -\dfrac{6}{7}. 
\end{equation}
We used the formula the Hubble's law of expansion to determine the ${\cal H}_1$ integration constant and the numerical integration for the rotating case,
\begin{equation}
    \dfrac{\Dot{a}(t)}{a(t)} \bigg|_{t=t_0} = H_0, \text{ if} \quad a(t_0) = 1 \ ,
\end{equation}
where $H_0 = 66.6^{+4.1}_{-3.3}$ km/s/Mpc is the experimental value of the Hubble-constant \cite{ref1-Hubble}. The expansion rates for the rotating (yellow line) and non-rotating (blue line) Universe are drawn on figure~\ref{fig::HubbleTimeEvolution} in the usual units of the standard cosmology. The dashed line represents the measured value of the Hubble parameter today. 
\newpage
\subsection{Connection to the critical density}

Energy conservation has to be investigated as well, which leads to the determination of the critical density parameter of the Universe. We apply the Newton equation to obtain the so-called acceleration equation~\cite{ref-Newtonian}
\begin{equation}
    \dfrac{\Ddot{a}(t)}{a(t)} = - \dfrac{4\pi G}{3} \rho(R(t),t) \quad \Rightarrow \quad \Ddot{R}(t) = - \dfrac{4\pi G}{3} \rho(R(t),t) R(t) \ . 
    \label{eq:acc}
\end{equation}
Let us multiply the equation with $\Dot{R}(t)$ and integrate over time and the Eq.~\eqref{eq:acc} becomes
\begin{equation} \label{eq::energycons}
    \Dot{R}(t)^2 = \dfrac{8\pi G}{3} \rho(t) R^2(t) + U_0,  
\end{equation}
which is the energy conservation equation.  In general relativity, the $U_0$ is related to the curvature of space-time. Still, since we are working in the framework of Newtonian cosmology, it is associated with a kind of mechanical energy. The equation~\eqref{eq::energycons} can be transformed into the following form if we use the $H(t):= \Dot{R}(t)/R(t)$ definition
\begin{equation}
    H^2(t) = \dfrac{8\pi G}{3} \rho(R(t),t) + U_t,  
\end{equation}
where $U_t = U_0/R(t)^2$ is a dynamical constant. The $U_0 = H_0^2 (1 - \Omega_{0,k})$ and we used the definition of the critical density 
\begin{equation}
    \rho_{0} = \Omega_0 \rho_c, \text{ where} \quad \rho_c = \dfrac{3H_0^2}{8\pi G}    \ . 
\end{equation}
The value $U_t$ dynamical constant determines the evolution of the system. On one hand, if the $U_t > 0$, then the Universe will eventually re-collapse. On the other hand, if $U_t < 0$ the system will expand to infinity. The $U_t = 0$ is a special case when the Universe shows similar expanding behavior but it will reach zero velocity in infinite time~\cite{ref-Newtonian}. We are searching for an isentropic solution therefore, we imply the entropy conservation $\mathrm{d}E + p \mathrm{d}V = 0$ equation from Ref.~\cite{ref-Ryden}. If we use the non-steady flow internal energy formula $E = V \rho (u+1)$ in $c=1$ unit system and use the equation of state from Eq.(3) in Ref.~\cite{ref-Landau}. The entropy conservation~\cite{ref-cosmology1} can be expressed in terms of 
\begin{equation}
    \Dot{E} + p\Dot{V} = 0 \quad \Rightarrow \quad  \dfrac{\Dot{\rho}}{\rho} = - 3 \left[ \dfrac{\Dot{R}}{\Dot{R}} + \dfrac{\Ddot{R}} {\Dot{R}} \right],
\end{equation}
with the solution of
\begin{equation}
    \dfrac{\rho}{\rho_0} = H_0^3\left( \Dot{a}(t) a(t) \right)^{-3}.
\end{equation}
Thus, the relationship between the time-dependent Hubble parameter and the density parameter in this model is
\begin{equation}
    H(t) = H_0^2\sqrt{\Omega_{0,CDM} \left( \frac{H_0}{H(t)}\right)^{3} \frac{1}{a^{3}(t)}+ \Omega_{DE,0} \frac{1}{a^{2}(t)}}. 
\end{equation}
which cannot be solved explicitly. The $\Omega_{0, CDM}$ or $\Omega_{DE,0}$ are density parameters related to the cold dark-matter (CDM) and dark energy (DE) respectively.  The relation between the density parameters and the dynamical parameters of the self-similar solution is defined and explained in the next section.
\section{Discussion}

According to current scientific understanding, dark matter and dark energy make up about 95\% of the total energy density of the observable Universe today. The dark fluid theory suggests that a single substance may explain dark matter and dark energy. The behaviour of the hypothetical dark fluid is believed to resemble that of cold dark matter on galactic scales while exhibiting similar characteristics to dark energy at larger scales~\cite{ref-Farnes}. Predictions can be obtained from our Sedov\,–\,von Neumann\,--\,Taylor blast wave-inspired non-relativistic dark fluid model on galactic and cosmological scales. A useful feature of the model is that the initial value problem of the reduced ordinary differential equation system is easier to handle than the boundary- and the initial condition problem of the original partial differential equations. To provide a reliable practical basis for our dark-fluid model, we test the theoretical results on astrophysical scales, presenting the similar nature of the solution on the cosmological scale.

The solution was developed based on cosmological observations, using the Hubble law to scale the expansion of the Universe. Our model includes various scaling mechanisms through the use of a Sedov-type self-similar \emph{ansatz}, which allows for describing different time decay scenarios \cite{ref-Hubble}. In the case of the non-rotating system, we can conclude that the radial velocity profile of the solution provides Hubble-like expansion. However, the non-rotating model provides inflation-like behavior, $u(r,t) > 1$, on a long-range timescale, which cannot be physical (causal). We have also seen that the high initial velocity of the dark fluid will relax to a small, constant non-relativistic value at the long timescale ($t > 6$ billion years). Also, an interesting feature of the model is that the gravitation potential is a negative polynomial in time, which is consistent with the distribution of the density of the dark fluid.

One notable aspect in the case of the rotating model is that it does not show super-luminous behavior in the expected time range, contrary to the non-rotating model. Simultaneously, we have found that the radial profile of the density will saturate and becomes close to flat at far distances from the initial point (see top right panel of Fig.~\ref{fig6}). One may also set the initial condition according to that the Universe today is observed as flat Euclidean, with the density parameter
\begin{equation}
\Omega(t) = \dfrac{\sum_i\rho_i}{\rho_c} \approx 1,   
\end{equation}
is the corresponding critical density with Hubble-parameter, $H_0$. The flatness of the Universe is indicated by the recent measurements of WMAP \cite{ref-WMAP}. The sum index stands for baryonic $(B)$, dark energy (DE), and cold-dark matter (CDM) respectively. We can define the matter part of the $\Omega_{M} = \Omega_{CDM} + \Omega_{B}$, and the full $\Omega = \sum_i \Omega_i$ and $i \in \{B,CDM,DE\}$ \cite{ref-Valev}. If the $\Omega_{CDM} \gg \Omega_B$ relation is correct, then we could assume the following identity,
\begin{equation} \label{eq::energyration}
    \dfrac{\Omega_M(t)}{\Omega(t)} \bigg|_{t=t_0} \sim \dfrac{E_{kin}(R(t),t)}{E_{tot}(R(t),t)} \bigg|_{t=t_0} = 0.26. 
\end{equation}
Accordingly, we can determine the relevant time and radial coordinates from the obtained results which correspond to this specific energy relation is evaluated at $r = R(t_0)$ and $t = t_0$. In the absence of dark fluid, the Universe will continue to expand indefinitely, but at a gradually slowing rate that will eventually approach zero. This will cause an open topology universe. The ultimate fate of the Universe is that the temperature asymptotically approaches absolute zero, the so-called "Big Freeze". At the same time, it is  essential to mention, that our non-relativistic model weakness does not provide as precise results as relativistic Friedmann-equation-based models~\cite{ref-Farnes}. 

\section{Conclusions and Outlook}

In this paper, we studied the behaviour of self-similar time-dependent solutions in a coupled non-linear partial differential equation system describing a non-viscous, non-relativistic, and self-gravitating fluid (Euler-Poisson system). The reason behind the applied self-similar solutions is that they are proven to be a very efficient method to analyze various kinds of physical systems. Especially to analyze the hydrodynamical description of systems that involves collapse and explosion. 

The analysis presented in this paper is an Euler-Poisson extension of our previous model~\cite{ref-Pocsai}. We have found that Sedov-Taylor type of solutions exists, and the algebraic equation obtained for the similarity exponents has only one unique solution. 
We have used the obtained solution to describe the behaviour of the non-relativistic dark fluid on cosmological scales, and we presented the relevant kinematical and dynamical quantities. We also showed that our model based on self-similarity is in agreement with the Newtonian Friedmann equations.

Although, one can easily notice that the model has certain limitations, due to its classical nature. Yet, it does provide relatively adequate results on cosmological scale. We showed that the obtained quasi-analytical solution for the evolution of the Hubble parameter is in agreement with the standard cosmological model e.g.~\cite{ref2-Hubble}. Also, we presented in the previous section that the energy ratio from our model is resembling with the ratio of $\Omega_M (t)/\Omega (t)$ density parameters as of today.

It has the practical benefit, that the calculation does not need high computing performance and resources. Therefore, it could be used to estimate the physical value of the initial- and boundary values, when more sophisticated theoretical or numerical simulations are used. Moreover, it can provide a reliable basis for comparison for 2- or 3-dimensional hydrodynamical simulations. Also, it is possible to improve this model with reasonable effort to describe even relativistic matter~\cite{ref-Rezzola}, Chaplygin gas~\cite{ref-Chaplygin}, or two-fluid models.

%
\vspace{6pt} 
\authorcontributions{
Original idea: I.F. B., Formal analysis, B.E.Sz.; Software, B.E.Sz.; Visualization, B.E.Sz.; Writing—original draft, B.E.Sz., Correction I.F.B, and G.G.B.  All authors have read and agreed to the published version of the manuscript.}
\funding{Authors gratefully acknowledge the financial support by the Hungarian National Research, Development and Innovation Office (NKFIH) under Contracts No. OTKA K135515, No. NKFIH 2019-2.1.11-TET-2019-00078, and No. 2019-2.1.11-TET-2019-00050 and Wigner Scientific Computing Laboratory (WSCLAB, the former Wigner GPU Laboratory).}

\dataavailability{This work is based on an analytic calculation of the given formulae. All data are the plots. The data underlying this article will be shared on reasonable request to the corresponding author.} 

\acknowledgments{Authors gratefully acknowledge the useful discussions with N. Barankai. One of the authors (IFB) offers this study in memory of an astronomer György Paál (1934 - 1992) who taught him physics and sailing in the summer of 1990 at Lake Balaton.
}

\conflictsofinterest{The authors declare no conflict of interest.}

\appendixtitles{no} 


\begin{adjustwidth}{-\extralength}{0cm}

\reftitle{References}

\appendix

\section*{Appendix A}
\label{app:a}

Here, in table~\ref{tab:unit} we summarize the used macroscopical and astronomical scales, quantities, and couplings in different units: SI and geometrized. We also provide the factors to convert them. Also, we included astronomical units~\ref {tab:unit2} which are used in the Discussion section. 
\begin{table}[H]
    \centering
    \begin{tabular}{ l c c c}
        \hline
         Variable & SI Unit & Geom Unit. & Factor  \\
         \hline
         \hline
         mass & kg & m & $c^2G^{-1}$\\
         length & m & m & 1 \\
         time & s & m & $c^{-1}$ \\
         density & kg $m^{-1}$ & $m^{-2}$ & $c^2G$ \\
         velocity & m $s^{-1}$ & 1 & $c$ \\
         acceleration &  m $s^{-2}$ & m$^{-1}$ & $c^2$ \\
         force & kg m $s^{-2}$ & 1 & $c^4 G^{-1}$\\
         energy & kg m$^2$ $s^{-2}$ & m$^{-1}$  & $c^4 G^{-1}$\\
         energy density & kg m$^{-1}$ $s^{-2}$ & m$^{-2}$ &  $c^4 G^{-1}$\\
         \hline
\end{tabular}
    \caption{The relevant physical quantities in SI and in geometrized units. To convert geometrized units into SI, one should use the factors.}
    \label{tab:unit}
\end{table}       
\begin{table}[H]
    \centering
    \begin{tabular}{ l c c}
         \hline
         Variable & Astronomical Unit. &  SI Unit  \\
         \hline
         \hline
         length &  ly & $9.46073047258\cdot10^{15}$m\\
         length &  Gly & $9.46073047258\cdot10^{24}$m\\
         length &  kPc & $3.08567758128\cdot10^{19}$m\\
         time & Gy & $3.1556926 \cdot 10^{16}$s\\
         \hline
    \end{tabular}
    \caption{The relevant astronomical quantities and the corresponding value in SI. }
    \label{tab:unit2}
\end{table}

\end{adjustwidth}
\end{document}